\documentclass[%
reprint,
superscriptaddress,
 amsmath,
 amssymb,
 aps,
]{revtex4-2}

\usepackage{graphicx}
\usepackage{dcolumn}
\usepackage{bm}
\usepackage{hyperref}
\usepackage{color}

\begin{document}

\preprint{APS/123-QED}

\title{Strong Coupling from Inclusive Semi-leptonic Decay of Charmed Mesons}

\author{Jinfei Wu}%
\affiliation{%
 \textit Institute of High Energy Physics, Chinese Academy of Sciences, Beijing, 100049, Beijing, China
}%
\affiliation{%
 \textit China Center of Advanced Science and Technology, Beijing, 100190, Beijing, China
}%
\author{Xinchou Lou}
 \affiliation{%
 \textit Institute of High Energy Physics, Chinese Academy of Sciences, Beijing, 100049, Beijing, China
}%
\affiliation{%
 \textit University of Texas at Dallas, Richardson, 75083, Texas, USA
}%
\affiliation{%
 \textit Center for High Energy Physics, Henan Academy of Sciences, Zhengzhou, 450046, Henan, China
}%
\author{Yuzhi Che}
 \affiliation{%
 \textit Institute of High Energy Physics, Chinese Academy of Sciences, Beijing, 100049, Beijing, China
}%
\affiliation{%
 \textit University of Chinese Academy of Sciences, Beijing, 100049, Beijing, China
}
\author{Gang Li}
 \affiliation{%
 \textit Institute of High Energy Physics, Chinese Academy of Sciences, Beijing, 100049, Beijing, China
}%
\author{Yanping Huang}
 \affiliation{%
 \textit Institute of High Energy Physics, Chinese Academy of Sciences, Beijing, 100049, Beijing, China
}%
\author{Manqi Ruan}
 \affiliation{%
 \textit Institute of High Energy Physics, Chinese Academy of Sciences, Beijing, 100049, Beijing, China
}%
\author{Jingbo Ye}
 \affiliation{%
 \textit Institute of High Energy Physics, Chinese Academy of Sciences, Beijing, 100049, Beijing, China
}%

\date{\today}

\begin{abstract}
Employing the heavy quark expansion model with the kinetic scheme, we evaluate $\alpha_S(m_c^2)$, the strong coupling constant at the charm quark mass $m_c$ with data on inclusive semileptonic decays of charmed mesons.
Using the experimental values of semileptonic decay widths of the $D^0$ and the $D^+$, the value of $\alpha_{s}(m_c^{2})$ is determined to be $0.445\pm0.009\pm0.114$, where the first uncertainty is experimental and the second systematic.
This reported $\alpha_{s}(m_c^{2})$ is in good agreement with the value of $\alpha_{s}(m_c^{2})$
calculated by running $\alpha_S(m_Z^2)$ at the $Z^0$ boson mass $m_Z$ with the renormalization group evolution equation. In addition, values of $\alpha_{s}(m_c^{2})$ obtained individually from each of the $D^0$, $D^+$, and $D_s^+$ mesons are found to be consistent being of the same origin.
 
\end{abstract}

\maketitle


\section{I. Introduction}
\label{sec:intro}

In the Standard Model of elementary particle physics, Quantum Chromo-Dynamics (QCD) is the gauge field theory for the strong interaction. 
In QCD, the gluons are force mediators and $\alpha_S$, the effective strong coupling constant, dictates many features of the strong interaction.
The asymptotic freedom is one of the important features of QCD, where the strength of $\alpha_S$ increases with decreasing energy scale.
The value of $\alpha_S$ has been measured over the energy scale  ranging from the $\tau$ lepton mass $m_\tau$, to several $\mathrm{TeV}$, and is found to be consistent with the theoretical prediction.
However, $\alpha_S$ at energies below $m_\tau$ has not been measured, and the QCD physics in this region may enter the non-perturbative scheme and exhibit unknown behaviors. It is very desirable to measure $\alpha_S$ at lower energies to further understand QCD and probe possible new physics. 

Significant progress has been made in the theoretical description of inclusive semileptonic decays of charmed and B mesons using the framework of heavy quark expansion model (HQE) in the past decades \cite{Ellis:1975hr, Isgur:1988gb, Bigi:1993fe, Blok:1993va, Gambino:2010jz, Fael:2019umf, Fael:2020tow}.
In the HQE framework, the features of the inclusive semileptonic decays of heavy quarks are expressed in terms of $\alpha_S$, quark masses, CKM matrix elements, and non-perturbative parameters.
HQE calculations describe well experimental features of inclusive semileptonic decays of charmed and B mesons \cite{Gambino:2010jz, Fael:2019umf, Fael:2020tow, Buchmuller:2005zv, King:2021xqp, Finauri:2023kte}.
The HQE has been employed as a reliable method to extract the b quark mass and $|V_{cb}|$ experimentally with inclusive semileptonic decays of B mesons \cite{Buchmuller:2005zv, Finauri:2023kte, Gambino:2011cq, Gambino:2013rza, Alberti:2014yda, Bernlochner:2022ucr}.
In these studies, the b quark mass and $|V_{cb}|$ are determined from the fits to the observables of inclusive semileptonic decays of B mesons, where $\alpha_S$ is fixed to the value running from $\alpha_S(m_Z^2)$. 

Similarly, the procedure can be applied to inclusive semileptonic decays of charmed mesons. 
Experimentally, measurements of $m_c$ and $|V_{cs}|$ have become more precise \cite{ParticleDataGroup:2022pth, Fael:2020iea, Fael:2020njb} which will make it possible to 
determine $\alpha_S(m_c^2)$ as a parameter from charmed mesons, either by fixing the values of $m_c$ and $|V_{cs}|$ to those measured in processes other than semileptonic D decays, or through a fit that will extract $m_c$, $|V_{cs}|$ and $\alpha_S(m_c^2)$ simultaneously from inclusive semileptonic decays of charmed mesons.
In this article we will present a determination of $\alpha_S(m_c^2)$ from inclusive semileptonic decays of charmed mesons.

\section{II. Heavy quark expansion model in kinetic scheme}
\label{sec:HQE}

The theoretical calculation \cite{Gambino:2010jz} of the inclusive semileptonic decay width ($\Gamma_{SL}$) for charmed mesons is employed to extract $\alpha_S(m_c^2)$ in this study.
In \cite{Gambino:2010jz}, the authors considered $\mathcal{O}(\alpha_S)$ and $\mathcal{O}(\beta_0\alpha_S^2)$ corrections \cite{Czarnecki:1994pu, Aquila:2005hq}, as well as $\mathcal{O}(1/m_c^3)$ contributions \cite{Gremm:1996df} in calculating $\Gamma_{SL}$.
As shown in Eq.~\ref{eq:gsl} \cite{Gambino:2010jz}, $\Gamma_{SL}$ is expressed in terms of $\alpha_S(m_c^2)$, quark masses, CKM matrix element $|V_{cs}|$, and non-perturbative corrections.
In Eq.~\ref{eq:gsl}, $G_F$ is the Fermi coupling constant, $r$ is the square of ratio between strange and charm quark mass ($m_s^2 / m_c^2$), $\alpha_S\equiv\alpha_S(m_c^2)$, $\mu^2_{\pi}$ and $\mu^2_{G}$ are the kinetic and the chromomagnetic dimension-five operators \cite{Bigi:1992su, Blok:1992he, Bigi:1997fj}, 
$\rho_{D}^3$ and $\rho_{LS}^3$ are the Darwin and the spin-orbital (LS) dimension six operators \cite{Bigi:1997fj} in HQE model.
The weak annihilation (WA) contribution, $B_{WA}$,  depends on the type of the spectator quark within each charmed meson.
\begin{equation}
    \begin{aligned}
        \Gamma_{SL} = & \frac{G_F^2 m_c^5}{192\pi^3} |V_{cs}|^2 [f_0(r) + \frac{\alpha_S}{\pi}f_{1}(r) + \frac{\alpha_S^2}{\pi^2}f_2(r)\\
        & + \frac{\mu_{\pi}^{2}}{m_c^2} f_{\pi}(r) + \frac{\mu_{G}^{2}}{m_c^2} f_{G}(r)
        + \frac{\rho_{LS}^{3}}{m_c^3} f_{LS}(r) \\
        & + \frac{\rho_{D}^{3}}{m_c^3} f_{D}(r) + \frac{32\pi^{2}}{m_c^3} B_{WA}]
    \end{aligned}
    \label{eq:gsl}
\end{equation}
The coefficients of perturbative and non-perturbative items, $f_{0,1,2}(r)$ and $f_{\pi, G, LS, D}(r)$, are calculated in Eq.~\ref{eq:f} \cite{Gambino:2010jz},
where $n_f$ is the number of active flavors and $\beta_0$ is the QCD beta function, $\beta_0 = 11 - 2n_f/3$.
\begin{equation}
	\begin{aligned}
		f_0(r) & = 1 - 8r + 8r^3 - r^4 - 12r^2\cdot \mathrm{log}(r), \\
		f_1(r) & = 2.86\sqrt{r} - 3.84r\cdot \mathrm{log}(r), \\
		f_2(r) & = \beta_0[8.16\sqrt{r} - 1.21r\cdot \mathrm{log}(r) - 3.38], \\
		f_{\pi}(r) & = - f_0(r) / 2 ,\\
		f_{G}(r) & = \frac{1}{2}f_0(r) - 2(1-r)^4, \\
		f_{LS}(r) & = - f_{G}(r), \\
		f_{D}(r) & = \frac{77}{6} + \mathcal{O}(r) + 8 \mathrm{log}(\frac{\mu_{WA}^2}{m_c^2}), \\
	\end{aligned}
    \label{eq:f}
\end{equation}
The infrared cutoff scale $\mu$ in the kinetic scheme is set to be $0.5\ \mathrm{GeV}$ in this study.
In the theoretical expression of $f_D(r)$, $0.8\ \mathrm{GeV}$ is treated as the $\overline{\mathrm{MS}}$ renormalization scale ($\mu_{WA}$) associated to mix of the Darwin and WA operators \cite{Gambino:2010jz, Gambino:2005tp, Gambino:2007rp}.
In Eq.~\ref{eq:gsl}, only the process of $c\rightarrow s l\bar{\nu}$ is taken into account, which is slightly different from experimental measurements \cite{CLEO:2009uah, BESIII:2021duu} due to missing Cabibbo suppressed processes.
A corresponding systematical uncertainty is assigned to cover the missing processes in the determination of $\alpha_S(m_c^2)$.

\section{III. Fit method}
\label{sec:fit}

A $\chi^2$ minimization method is employed to determine $\alpha_S(m_c^2)$ from fits to $\hat{\Gamma}_{SL}$, which is the $\Gamma_{SL}$ expression of Eq.~\ref{eq:gsl} for different charmed mesons. The $\chi^2$ function is constructed as  

\begin{equation}
	\begin{aligned}
		\chi^2(\alpha_S, \theta_{j}) & = \sum_i \frac{[\Gamma_{SL, D_i} - \hat{\Gamma}_{SL}(\alpha_S, \theta_{j})]^2}{\sigma_{\Gamma_{SL, D_i}}^2} + \sum_j \frac{(\theta_j - \theta^{\prime}_j)^2}{\sigma_{\theta^{\prime}_j}^2}  , \\
	\end{aligned}
    \label{eq:chi2}
\end{equation}
where $D_i$ denotes $D^{+}$, $D^{0}$, and $D^{+}_s$ respectively,
$\Gamma_{SL, D_i}$ and $\sigma_{\Gamma_{SL, D_i}}$ are measured inclusive semileptonic decay width and corresponding uncertainty of $D_i$.
In Eq.~\ref{eq:chi2}, $\theta_j =$ $\{m_c$, $m_s$, $|V_{cs}|$, $\mu_{G}^2$, $\mu_{\pi}^2$, $\rho_{D}^3$, $\rho_{LS}^3\}$ represents constrained parameters, whose values and uncertainties are $\theta^{\prime}_j$ and $\sigma_{\theta^{\prime}_j}$, respectively.

In this study, $G_{F}$ is fixed to $1.1663788\times 10^{-5}$ \cite{ParticleDataGroup:2022pth}.
According to studies in \cite{Gambino:2010jz}, the values of $B_{WA}$ for $D^{+}$, $D^0$, and $D^{+}_s$ are fixed at $-0.001$, $-0.001$, and $-0.002\ \mathrm{GeV}^3$, respectively.
Except for $G_F$ and $B_{WA}$, other parameters are allowed to float in the determination of $\alpha_S(m_c^2)$.
The $|V_{cs}|$ has been measured to be $0.975\pm0.006$ \cite{ParticleDataGroup:2022pth}.
In the kinetic scheme, expected values of $\mu_{G}^2$ and $\rho_{LS}^3$ do not run with respect to energy scale, which have been determined to be $0.288\pm0.049\ \mathrm{GeV}^2$ and $-0.113\pm0.090\ \mathrm{GeV}^3$ from inclusive semileptonic B decays \cite{Finauri:2023kte}.
In \cite{Gambino:2010jz, Buchmuller:2005zv, Gambino:2004qm}, values of $\mu_{\pi}^2(0.5\ \mathrm{GeV})$ and $\rho_{D}^3(0.5\ \mathrm{GeV})$ have been determined to be $0.26\pm0.06\ \mathrm{GeV}^2$ and $0.05 \pm 0.04\ \mathrm{GeV}^3$, that are evolved to $\mu=0.5\ \mathrm{GeV}$ using $\mathcal{O}(\alpha_S^2)$ expressions from values in $\mu=1\ \mathrm{GeV}$.
The mass of the strange quark is set to be $93.4\pm8.6\ \mathrm{MeV}$ \cite{ParticleDataGroup:2022pth}. 

The convergence of perturbative series of the $\Gamma_{SL}$ expression is strongly affected by the mass definition of the charm quark \cite{Beneke:1994sw, Bigi:1994em, Bigi:1996si, Melnikov:2000qh}.
In \cite{Fael:2020iea}, the pole mass and the $\overline{\mathrm{MS}}$ scheme exhibit bad convergence behaviours in QCD corrections to $\Gamma_{SL}$.
To avoid the divergence, the kinetic scheme \cite{Beneke:1994sw, Bigi:1996si, Antonelli:2009ws} is introduced, and is adopted to calculate $\Gamma_{SL}$.
The relationship between $\overline{\mathrm{MS}}$ and kinetic mass of charm quark has been investigated to three-loop order ($\mathrm{N}^3$LO) \cite{Fael:2020iea, Fael:2020njb}.
For different choices of $\mu_s$ ($\overline{\mathrm{MS}}$ scale), the value of $m_c$ at scale of $0.5\ \mathrm{GeV}$ in kinetic scheme $m_c^{kin}(0.5\ \mathrm{GeV})$ has been obtained separately using the relationship in \cite{Fael:2020iea, Fael:2020njb}, as shown in Eq.~\ref{eq:mc}.

\begin{equation}
	\begin{aligned}
		m_c^{kin}(0.5\ \mathrm{GeV}) & = 1336\ \mathrm{MeV}\ \mathrm{for}\  \overline{m}_c(\mu_s = 3\ \mathrm{GeV}) \\
		m_c^{kin}(0.5\ \mathrm{GeV}) & = 1372\ \mathrm{MeV}\ \mathrm{for}\  \overline{m}_c(\mu_s = 2\ \mathrm{GeV}) \\
		m_c^{kin}(0.5\ \mathrm{GeV}) & = 1404\ \mathrm{MeV}\ \mathrm{for}\  \overline{m}_c(\mu_s = \overline{m}_c) \\
	\end{aligned}
    \label{eq:mc}
\end{equation}
The average of $m_c^{kin}(0.5\ \mathrm{GeV})$ from different $\mu_s$ is treated as input value of $m_c(0.5\ \mathrm{GeV})$ in the $\chi^2$ fit, which is determined to be $1370\ \mathrm{MeV}$.
For a conservative consideration, the largest difference between $m_c(0.5\ \mathrm{GeV})$ and $m_c^{kin}(0.5\ \mathrm{GeV})$ is taken as an uncertainty on $m_c(0.5\ \mathrm{GeV})$.
To investigate the bias due to the choice of $m_c$ and $|V_{cs}|$, the first fit is performed with $m_c$ as a free parameter and $|V_{cs}|$ is allowed to vary within one standard error, and the second fit is done with $m_c$ and $|V_{cs}|$ both fixed at the world average. The results on $\alpha_S(m_c^2)$ from these fits are compared to check the consistency of the study.

\section{IV. Experimental inputs}
\label{sec:exp}

The experimental measurement of $\Gamma_{SL}$ is derived from the inclusive semileptonic decay branch fraction, $\mathcal{B}_{SL}$ \cite{CLEO:2009uah, BESIII:2021duu} and the lifetime, $\tau$ \cite{ParticleDataGroup:2022pth} with Eq.~\ref{eq:gsl_exp}, where $D_i$ denotes $D^{+}$, $D^{0}$, and $D^{+}_s$, respectively.
\begin{equation}
	\begin{aligned}
		\Gamma_{SL,\ D_i} & = \frac{6.582\times 10^{-25} \cdot \mathcal{B}_{SL}(D_i\rightarrow Xe\nu_{e})}{\tau_{D_i}}\ \mathrm{GeV} \\
	\end{aligned}
    \label{eq:gsl_exp}
\end{equation}
In Eq.~\ref{eq:gsl_exp}, $\tau_{D_i}$ is the mean life of $D_i$, and $\mathcal{B}_{SL}(D_i \rightarrow X e \nu_e)$ is the branch fraction of inclusive semileptonic decay for $D_i$.
The inclusive semileptonic branch fractions of $D^{+}$, $D^{0}$, and $D^{+}_{s}$ have been measured by the CLEO-c \cite{CLEO:2009uah} using $818\ \mathrm{pb}^{-1}$ and $602\ \mathrm{pb}^{-1}$ open-charm data at $E_{CM} = 3.774\ \mathrm{GeV}$ and $4.170\ \mathrm{GeV}$.
Because of limited statistic, the uncertainty of $\mathcal{B}_{SL, D^{+}_{s}}$ is much higher than $\mathcal{B}_{SL, D^{+}/D^{0}}$ in the measurements in CLEO-c.
Recently, $\mathcal{B}_{SL, D^{+}_{s}}$ has been also measured by the BESIII using $3.19\ \mathrm{fb}^{-1}$, $2.08\ \mathrm{fb}^{-1}$, and $1.05\ \mathrm{fb}^{-1}$ $e^+e^-$ collision data at $E_{CM} = 4.178\ \mathrm{GeV}$, $4.189 - 4.219\ \mathrm{GeV}$, and $4.225 - 4.230\ \mathrm{GeV}$ \cite{BESIII:2021duu}.
With more data, the uncertainty of $\mathcal{B}_{SL, D^{+}_{s}}$ has been reduced in the measurement by BESIII.
The $\mathcal{B}_{SL, D^{+}/D^{0}}$ from CLEO-c and $\mathcal{B}_{SL, D^{+}_{s}}$ from BESIII are adopted to calculate the $\Gamma_{SL}$ of $D^{+}$, $D^{0}$, and $D^{+}_{s}$.
In Tab.~\ref{tab:input_exp}, input values of $\mathcal{B}_{SL}(D_i \rightarrow X e \nu_e)$, $\tau_{D_i}$, and $\Gamma_{SL, D_i}$ are summarized.
The consistent $\Gamma_{SL}$ of $D^0$ and $D^+$ indicates reliability of HQE model in inclusive semileptonic decays of $D^0$ and $D^+$.
\begin{table}[htb]
    \centering
    \caption{The input values of $\mathcal{B}_{SL}(D_i \rightarrow X e \nu_e)$, $\tau_{D_i}$, and $\Gamma_{SL, D_i}$.}
    \label{tab:input_exp}
    \begin{ruledtabular}
    \begin{tabular}{cccc}
        $D_i$ & $\mathcal{B}_{SL}$ [\%] & $\tau$ [$10^{-13}$s] & $\Gamma_{SL}$ [$10^{-15} \mathrm{GeV}$]\\
        \hline
        $D^0$          & $6.46\pm0.09\pm0.11$ & $4.10\pm0.01$ & $104\pm2$\\
        $D^+$          & $16.13\pm0.10\pm0.29$ & $10.33\pm0.05$ & $103\pm2$\\
        $D^+_s$        & $6.30\pm0.13\pm0.10$ & $5.04\pm0.04$ & $82\pm2$\\
    \end{tabular}
    \end{ruledtabular}
\end{table}

\begin{figure}[htb]
    \centering
    \includegraphics[width=0.5\textwidth]{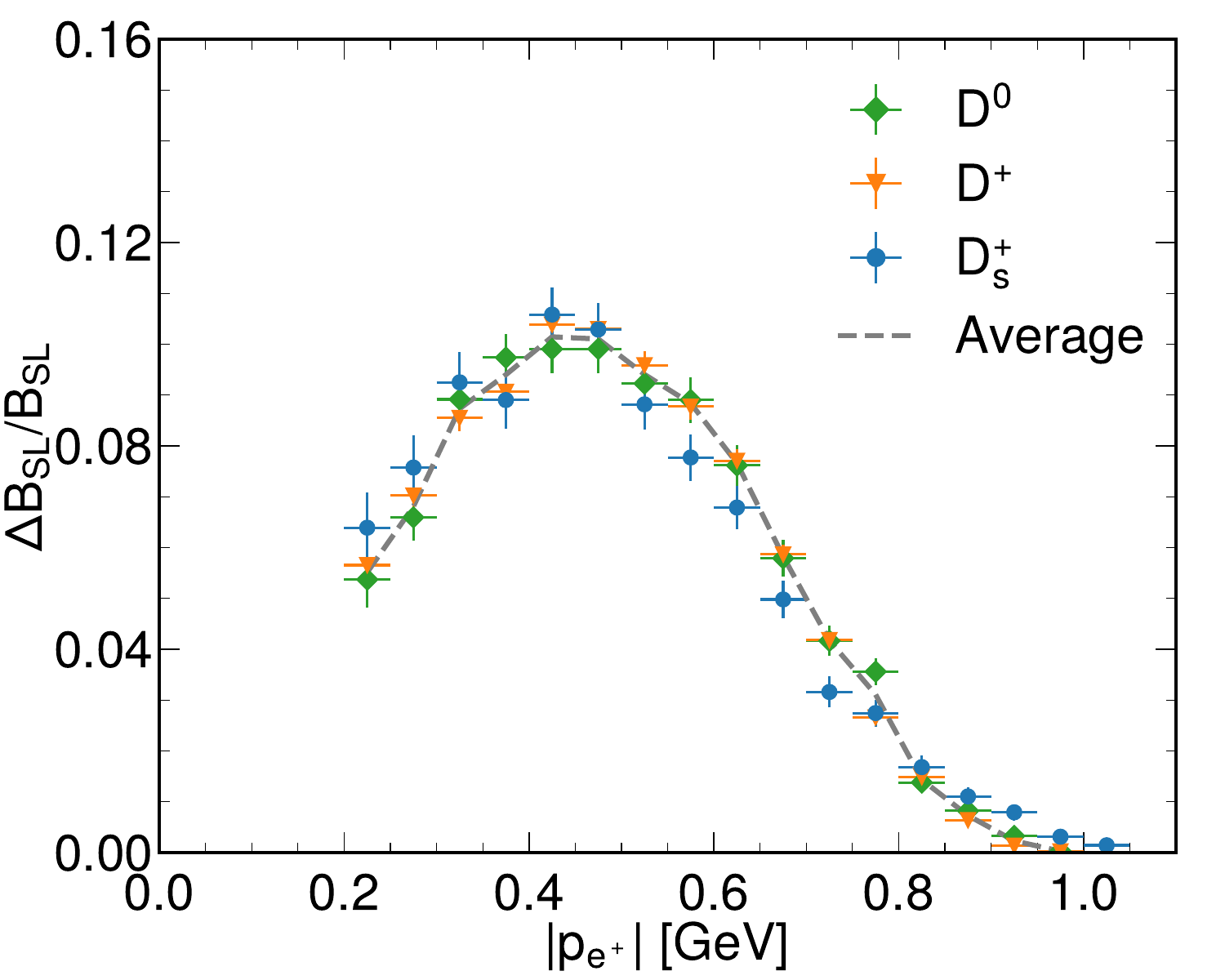}
    \caption{ The distributions of $|p_{e^{+}}|$ with $|p_{e^{+}}| > 200\ \mathrm{MeV}$ from inclusive semileptonic decays of $D^{+}$, $D^{0}$, and $D_s^+$ in laboratory frame. Green diamonds and orange triangles are respectively results of $D^0$ and $D^+$ measured by CLEO-c \cite{CLEO:2009uah}. Blue dots are data of $D_s^+$ measured by BESIII \cite{BESIII:2021duu}. The gray dashed line is the average for $D^0$ and $D^+$.}
    \label{fig:pe}
\end{figure}

Except for $\mathcal{B}_{SL}$, distributions of electron momentum ($|p_{e^{+}}|$) in laboratory frame have been measured in inclusive semileptonic decays of $D^{+}$, $D^{0}$, and  $D_s^+$ by the CLEO-c and the BESIII \cite{CLEO:2009uah, BESIII:2021duu}, that are shown in Fig.~\ref{fig:pe}.
Besides, the average $|p_{e^{+}}|$ of $D^0$, $D^+$, and $D_s^+$ is also plotted in Fig.~\ref{fig:pe}.
Kolmogorov-Smirnov (KS) tests \cite{2020SciPy-NMeth} between distributions of $|p_{e^{+}}|$ and $\overline{|p_{e^{+}}|}$ are performed to further validate the reliability of HQE model in inclusive semileptonic decays of charmed mesons.
The results of KS tests are obtained in Tab.~\ref{tab:ks}.
\begin{table}[htb]
    \centering
    \caption{Results of KS tests are listed, where the null hypothesis is that the two tested distributions are identical.} 
    \label{tab:ks}
    \begin{ruledtabular}
    \begin{tabular}{ccc}
        Test Distributions & Test Statistic & P Value \\
        \\
        \hline
        $|p_{e^{+}, D^0}|$ and $\overline{|p_{e^{+}}|}$  & $0.125$ & $1.000$ \\
        $|p_{e^{+}, D^+}|$ and $\overline{|p_{e^{+}}|}$  & $0.125$ & $1.000$ \\
        $|p_{e^{+}, D^+_s}|$ and $\overline{|p_{e^{+}}|}$  & $0.132$ & $0.992$ \\
    \end{tabular}
    \end{ruledtabular}
\end{table}

The distributions of $|p_{e^{+}}|$ agree well among $D^{+}$, $D^{0}$, and $D_s^+$. This is a strong indication of reliability of the HQE model on inclusive semileptonic decays of charmed mesons.
Since experimental measurements of $|p_{e^{+}}|$ are not available in the center of mass frame of charmed mesons,
only $\Gamma_{SL}$ is used to extract $\alpha_S(m_c^2)$ in this report.

\section{V. Results}
\label{sec:res}

The $\alpha_S(m^2_c)$ is extracted from $D^+$, $D^0$, and $D^+_s$, including
\begin{itemize}
    \item $D^+$, $D^0$, and $D^+_s$, respectively.
    \item $D^+$ and $D^0$ combined.
\end{itemize}
In the $\chi^2$ fit, high order perturbative corrections are needed to be taken into account in inclusive semileptonic decays of charmed mesons.
The  $\alpha_S^3$  order  correction to $b\rightarrow c l \bar{\nu}$ has been determined to be less than 1\% in the kinetic scheme \cite{Fael:2020tow}.
For conservative consideration, 5\% of $\Gamma_{SL}$ is taken as high order perturbative corrections in inclusive semileptonic decays of charmed mesons. 
Furthermore, the theoretical calculation of $\Gamma_{SL}$ in Eq.~\ref{eq:gsl} is contribution of $c\rightarrow s l \bar{\nu}$, where Cabibbo suppressed processes are missed.
To cover missed Cabibbo suppressed processes, $|V_{cd}|^2 / (|V_{cd}|^2+|V_{cs}|^2) \approx 5\%$ is treated to be an uncertainty of the $\Gamma_{SL}$ expression.
In total, $10\%$ is taken as the theoretical uncertainty in calculation of $\Gamma_{SL}$ for more conservative consideration.
The input values of the dimension six HQE matrix elements are evolved from the results obtained in the inclusive semileptonic B decays at $\mu$ = 1 GeV.
The treatment of inputs of dimension six HQE matrix elements may lead potential impact on systematical uncertainties in this study, 
which can be improved by the more precise measurements about inclusive semileptonic decays in the charm sector. 
Despite the kinetic scheme is adopted to improve the convergence of perturbative series, the contribution of higher order corrections may be larger due to the slow converge behavior in the charm sector, which could introduce possible underestimation on the systematical uncertainties.
As well as more measurements in the charm sector, such as spectral moments, can benefit for the determination of higher order corrections to reduce the corresponding systematical uncertainty.
The author need to emphasize that high order perturbative corrections play an important role in this study, and advanced theoretical calculations of high order perturbative corrections are highly desirable.

\begin{figure*}[htbp]
    \includegraphics[width=1.0\textwidth]{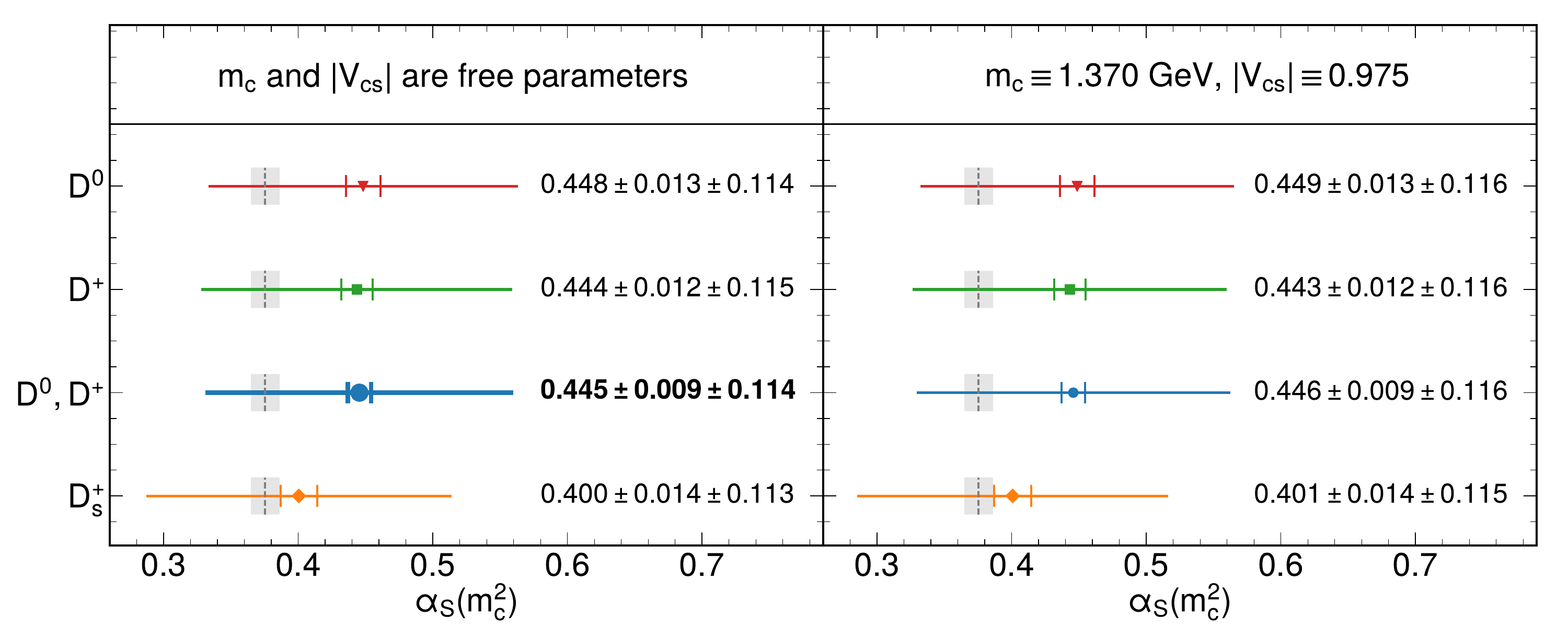}
    \caption{In the left panel, $m_c$ and $|V_{cs}|$ are allowed to float in the fit;  on the right panel  $m_c$ and $|V_{cs}|$ are fixed in the fit. Points with error bars are determined central values of $\alpha_S(m_c^2)$, and the inner and the outer error bars are experimental and total uncertainties, respectively.The grey dashed line and the box indicate the value of and the uncertainty on $\alpha_S(m_c^2)$ running to $m_c$ from $\alpha_S(m_Z^2)$.}
    \label{fig:alphaS}
\end{figure*}

\begin{figure}[htbp]
    \includegraphics[width=0.5\textwidth]{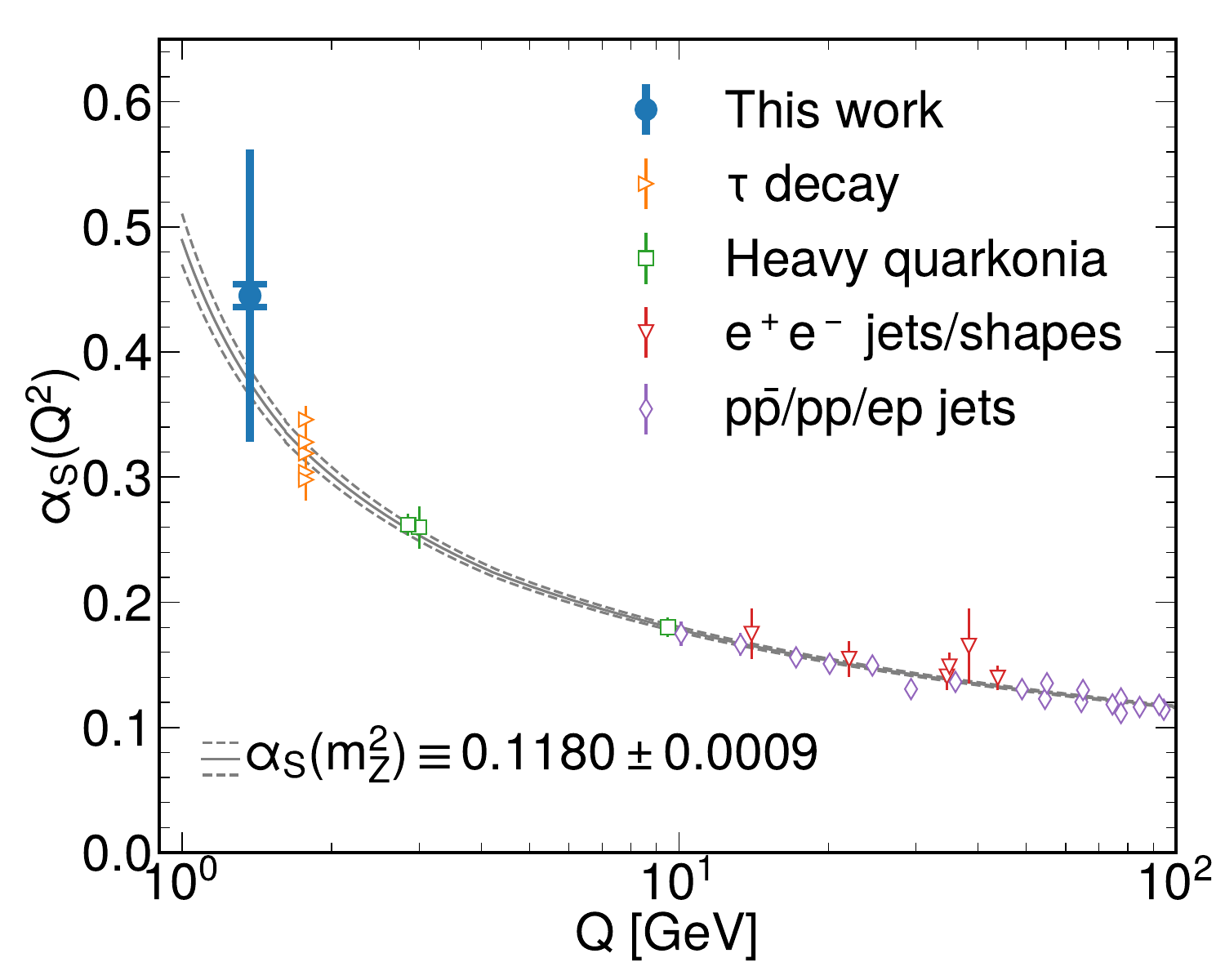}
    \caption{ Values of $\alpha_S$ at different energy scales. The blue dot is the measured $\alpha_S(m_c^2)$ in this study, where the inner and the outer error bars are experimental and total uncertainties, respectively. The other points are measurements of $\alpha_S$ at different energy scales \cite{Pich:2016bdg, Davier:2013sfa, Peset:2018ria, Boito:2014sta, Boito:2018yvl, Narison:2018dcr, Schieck:2012mp, D0:2009wsr, ZEUS:2012pcn, D0:2012xif, H1:2017bml}. The grey solid and the dashed lines are value and uncertainty of $\alpha_S$ running from $\alpha_S(m_Z^2)$, respectively. }
    \label{fig:runing}
\end{figure}

\begin{figure}[htbp]
    \includegraphics[width=0.5\textwidth]{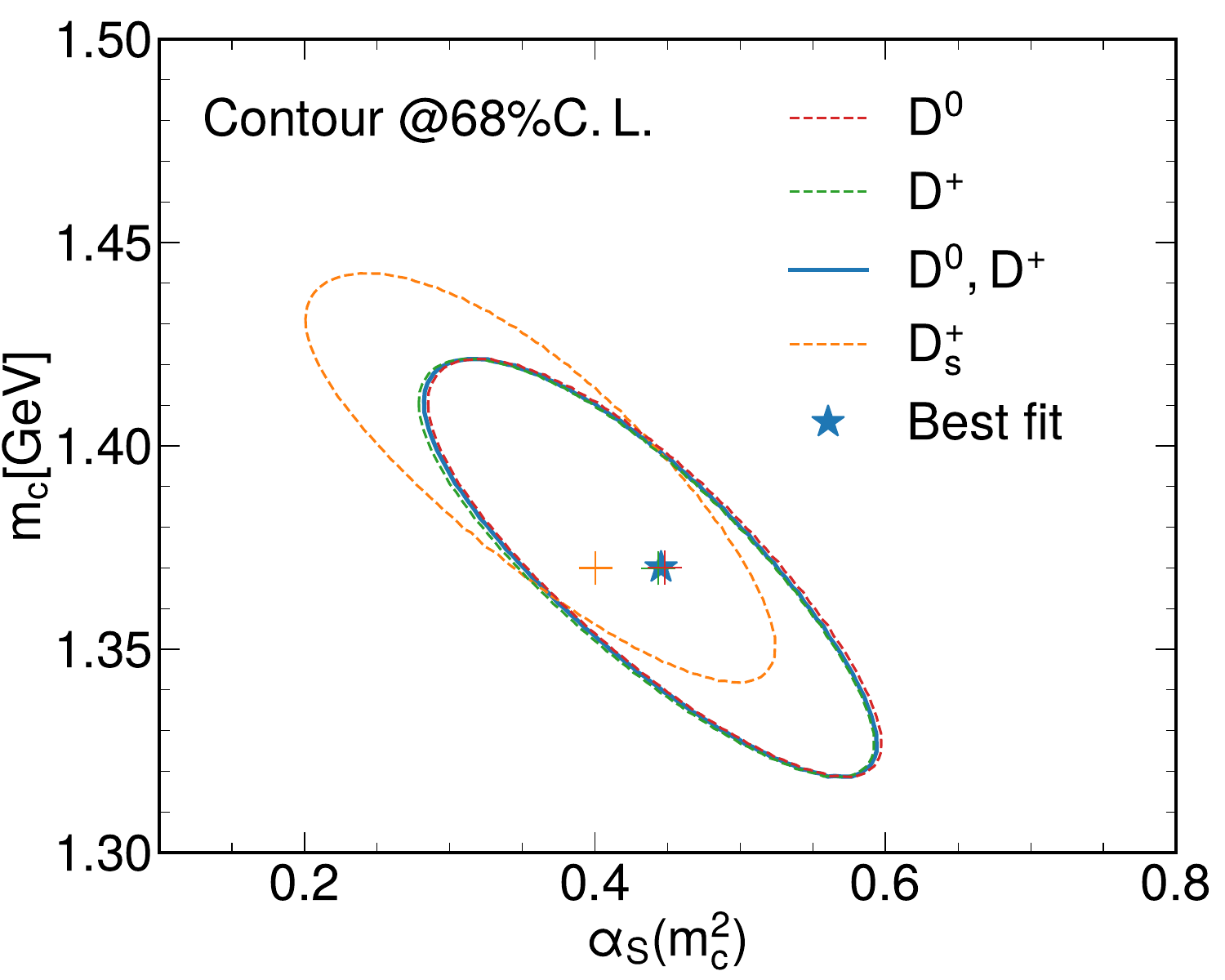}
    \caption{ The profile contours of different samples at 68\% confidence level. The solid blue curve and the star are the contour and the best fit value for $D^{0}$ and $D^{+}$ combined; the orange dashed curve and the cross are the contour and the best fit value for $D_{s}^{+}$; the red and the greed dashed curves and crosses are the contours and the best fit values for $D^{0}$ and $D^{+}$, respectively. }
    \label{fig:resd}
\end{figure}

In Fig.~\ref{fig:alphaS} and Tab.~\ref{tab:alphaS}, the fitted $\alpha_S(m^2_c)$ of each sample is shown and compared to $\alpha_S(m^2_c)$ running from $\alpha_S(m_Z^2)$ using $\mathbf{RunDec}$ \cite{Herren:2017osy} with renormalization group evolution equation.
\begin{table*}
    \caption{The value of $\alpha_S(m_c^2)$ is obtained for each sample, where values of $m_c$ and $|V_{cs}|$ are allowed to change in the fit. The first and second uncertainties of $\alpha_S(m_c^2)$ are experimental and theoretical, respectively. The reported result is obtained from $D^0$ and $D^+$ jointly (bold), which compares well with $0.375\pm0.011$, the value of $\alpha_S$ running from $m_Z$ down to $m_c$.}
    \label{tab:alphaS}
    \begin{ruledtabular}
    \begin{tabular}{ccccc}
        Sample & $D^0$ & $D^+$ & \boldmath{$D^{+}$, $D^{0}$} & $D^+_s$ \\
    \hline
        $m_c [\mathrm{GeV}]$ & $1.3701\pm0.0339$ & $1.3699\pm0.0340$ & \boldmath{$1.3701\pm0.0338$} & $1.3699\pm0.0340$ \\
        $\alpha_S(m_c^2) [10^{-3}]$ & $448\pm13\pm114$ & $444\pm12\pm115$ & \boldmath{$445\pm9\pm114$} & $400\pm14\pm113$ \\
    \end{tabular}
    \end{ruledtabular}
\end{table*}
Because of relative heavy spectator quark in $D_s^+$, the combined result of $D^+$ and $D^0$ is chosen to measure $\alpha_S(m_c^2)$ in this study. 
Using the sample of $D^0$ and $D^+$, $\alpha_S(m_c^2)$ is determined to be $0.445\pm0.009_{\mathrm{exp.}}\pm0.081_{m_c}\pm0.056_{\mathrm{trun.}}\pm0.057_{\mathrm{others}}$ at $m_c = 1.3701\ \mathrm{GeV}$,
where the first uncertainty is experimental, the second is due to the uncertainty on $m_c$, the third is associated with high order perturbative corrections in the $\Gamma_{SL}$ expression, and the fourth is related to other sources.
As shown in Fig.~\ref{fig:runing}, the measured value of $\alpha_S(m_c^2)$ is consistent within 1$\sigma$ with the value running from $\alpha_S(m^2_Z)$.
Among different charmed mesons, consistent values of $\alpha_S(m_c^2)$ indicate robustness of this method.
In the fit to the $D^{0}$ and $D^+$ sample, $\chi^2 / dof$ of the fit is $0.1/6$, indicating good agreement between data and the model.
In Fig.~\ref{fig:resd}, profile contours of different samples confirm good consistence among these charmed mesons and the robustness of this work.

The value of $m_c$ is fixed to $1.370\pm0.034\ \mathrm{GeV}$ to check stability of this study, and corresponding uncertainty is estimated by varying value of $m_c$ within $\pm1\sigma$.
Usually, the value of $|V_{cs}|$ is obtained from exclusive semileptonic or leptonic charmed meson decays, which may introduce bias in this study.
Hence a value of $|V_{cs}|$ without involving semileptonic charmed meson decays is necessary to validate this study.
With $|V_{cd}| = 0.2181\pm0.0049\pm0.0007$ from leptonic decays of $D^+$ \cite{HFLAV:2022esi} and $|V_{cb}| = (41.1\pm1.2)\times10^{-3}$ \cite{ParticleDataGroup:2022pth}, a value of $|V_{cs}|$ without involving semileptonic charmed meson decays is calculated to be $0.975\pm0.001$ by Eq.~\ref{eq:Vcs}, which has a negligible bias on the determination of $\alpha_S(m_c^2)$.
\begin{equation}
	\begin{aligned}
		|V_{cs}| = \sqrt{1 - |V_{cd}|^2 - |V_{cb}|^2} = 0.975\pm0.001 \\
	\end{aligned}
    \label{eq:Vcs}
\end{equation}
Fig.~\ref{fig:alphaS} presents fitted $\alpha_S(m_c^2)$ values for different D meson samples with fixed $m_c$.
The robustness of this study is confirmed by the consistent values of $\alpha_S(m_c^2)$ obtained by fits with fixed and floating $m_c$.

\section{VI. summary}
\label{sec:sum}

In summary, the value of $\alpha_S(m_c^2)$ at $m_c = 1.37\ \textrm{GeV}$ is determined to be $0.445\pm0.009\pm0.114$ using semileptonic decay widths of the $D^0$ and the $D^+$ measured by CLEO-c, and is cross-checked with $\Gamma_{SL}$ of the $D^+_s$ meson reported by BESIII. This result on  $\alpha_S(m_c^2)$  is in good agreement with the value obtained by running  $\alpha_S(m^2_Z)$ to $m_c$.
The values of $\alpha_S(m_c^2)$ are extracted for each of the $D^0$, $D^+$ and $D^+_s$ mesons and are found to be within $\pm 1\sigma$ among them, illustrating the robustness of analysis method used.
The leading uncertainty on $\alpha_S(m_c^2)$ is from the theoretical calculation of $\Gamma_{SL}$, which may be reduced with detailed experimental study of semileptonic decay of the D mesons and better HQE calculation.
This study represents the first measurement of $\alpha_S(m_c^2)$ with a new approach. 
With high statistics data and better modelling of the HQE, the systematic uncertainty with $\alpha_S(m_c^2)$ may be significantly reduced in future.

\section{acknowledgments}

The authors thank H. B. Li, X. T. Huang, X. Chen, G. Y. Zhang, J. L. Pei, H. Q. Zhang, Y. Q. Fang, and L. G. Shao for fruitful discussions.
This work is partially supported by the National Natural Science Foundation of China under grants No.12247119 and No.12042507.



\bibliography{apssamp}

\end{document}